\documentclass[a4paper,12pt]{article}

\def\II{I\!I}
\def\fifth{{\textstyle\frac{1}{5}}}

\usepackage{amsfonts}
\usepackage{amsmath,epsfig}
\numberwithin{equation}{section}
\begin{document}
\title{Small-amplitude inhomogeneous plane waves
     in a deformed Mooney--Rivlin material}

\author{Michel Destrade}
\date{2002}
\maketitle

\bigskip

\begin{abstract}
\noindent
The propagation of  small-amplitude inhomogeneous plane waves in an
isotropic homogeneous incompressible Mooney--Rivlin material is
considered when the material is maintained in a state of finite
static homogeneous deformation.
Disturbances of complex exponential type are sought and all
propagating inhomogeneous solutions to the equations of motion 
are given, as well as the conditions for linear, elliptical, 
or circular polarization.
It is seen that a great variety of solutions arises.
These include some original solutions, such as circularly-polarized
plane waves which propagate with an arbitrary complex scalar slowness,
or linearly-polarized waves for which the direction of propagation
is not necessarily orthogonal to the direction of attenuation.
Throughout the paper, geometrical interpretations and explicit
examples are presented.
\end{abstract}

\newpage


\section{Introduction}

The theoretical study of elastic plane waves has generated a large
literature in the area of finite elasticity. 
One type of plane waves which is of practical interest is that of 
motions propagating in a finitely and homogeneously deformed elastic 
material \cite{Biot40}, because the predeformation can be used to 
model, at least locally, the anisotropy of many mechanical and 
geological structures. 
A review of these topics can be found in a textbook by Ie\c{s}an 
\cite{Iesa89}.
Once it is homogeneously deformed, an elastic body presents three
privileged orthogonal directions, namely those of the principal axes of
the static deformation. 
Waves propagating in one of these directions are called `principal 
waves' \cite{Mana59} and their properties are well known and 
relatively easy to establish, because the principal directions offer 
a natural rectangular Cartesian coordinate system. 
Homogeneous plane waves propagating in an incompressible elastic 
material are necessarily transverse (the direction of propagation 
is orthogonal to the plane of polarization). 
Therefore, for such waves propagating in a nonprincipal direction, 
another rectangular Cartesian coordinate system proves to be useful, 
that formed by the direction of propagation and by two orthogonal 
directions in the plane of polarization. 
Currie and Hayes \cite{CuHa69} proved that the Mooney--Rivlin form 
for the strain energy density (which is used to model the mechanical 
behaviour of rubber \cite{Moon40, Trel75}), is the most general one 
for which homogeneous plane waves, be they of finite or of small 
amplitude, may propagate in \textit{any} direction for an arbitrary 
finite static homogeneous predeformation. 
Later, Boulanger and Hayes  \cite{BoHa92, BoHa95} studied in great 
detail finite-amplitude homogeneous plane waves in a deformed 
Mooney--Rivlin material.

Homogeneous plane waves are such that the planes of constant phase
(orthogonal to the direction of propagation) are parallel to the planes
of constant amplitude (orthogonal to the direction of eventual
attenuation). 
However, for a variety of physical problems, an attenuation of the 
amplitude occurs in a direction distinct from the direction of 
propagation. 
In those cases, a combination of `inhomogeneous' plane waves is 
introduced, usually in the form 
$e^{-\omega \mathbf{S'' \cdot x}}\cos \omega(\mathbf{S' \cdot x}-t)
 \mathbf{a}$, where $\omega$ is the frequency, and 
$\mathbf{S'}$, $\mathbf{S''}$, $\mathbf{a}$ are vectors. 
This procedure was successfully applied to various interfacial 
problems, such as reflected and refracted waves, Rayleigh
waves, Love waves, Stoneley waves, Scholte waves, etc. 
When the directions of propagation (that of the vector $\mathbf{S'}$),
of exponential attenuation (that of the vector $\mathbf{S''}$), and of
polarization (that of the vector $\mathbf{a}$) are orthogonal with
respect to one another, they form the basis for a rectangular 
Cartesian coordinate system in which the incremental equations 
of motion can be written and solved for small-amplitude 
\cite{Flav63,DoOg90} as well as for finite-amplitude \cite{Dest00} 
inhomogeneous waves in a deformed Mooney--Rivlin material. 
When these directions are not orthogonal, the equations of motion 
become much harder to solve. 
This is where the algebra of complex vectors,  also known as 
`bivectors', can play a very useful role. 
The  use of bivectors in order to describe inhomogeneous 
plane waves is made clear in a textbook by Boulanger and Hayes 
\cite{BoHa93}. 
Indeed bivectors, that is vectors with a real part and an imaginary 
part, can be used to describe the polarization of a wave through 
the `polarization bivector' $\mathbf{A}=\mathbf{A'}+i\mathbf{A''}$, 
as well as its propagation and attenuation through the `slowness 
bivector' $\mathbf{S}= \mathbf{S'}+i\mathbf{S''}$.
Hence an inhomogeneous plane wave is modelled as being proportional to
the real part of the expression 
$\mathbf{A} e^{i\omega (\mathbf{S\cdot x}-t)}$.

Previously, the propagation of small-amplitude \textit{inhomogeneous\/}
plane waves in a Mooney--Rivlin material subjected to a finite static
\textit{biaxial} homogeneous deformation has been considered first by
Belward \cite{Belw73}, and later by Boulanger and Hayes \cite{BoHa96},
using bivectors. 
In this paper, we consider the propagation of infinitesimal 
inhomogeneous plane waves in a Mooney--Rivlin material
which is maintained in a state of arbitrary \textit{triaxial} finite
static homogeneous deformation. 
Most results obtained here are a generalization and an extension 
to the case of inhomogeneous waves of results established by  
Boulanger and Hayes \cite{BoHa92} for the propagation of 
finite-amplitude \textit{homogeneous} plane waves in a
homogeneously deformed Mooney--Rivlin material. 
However, this extension is only possible when the amplitude 
of the wave is considered small enough to allow linearization. 
This restriction is due to the fact that finite-amplitude 
inhomogeneous plane waves can propagate in an incompressible 
elastic material only when they are linearly-polarized
\cite{Dest99}. 
For infinitesimal inhomogeneous plane waves, no such restriction 
applies and elliptical polarization is possible. 
The purpose of this paper is to find all inhomogeneous small-amplitude
plane waves of complex exponential type travelling in a deformed
Mooney--Rivlin material, and to establish the conditions for linear,
elliptical, and circular polarization.
Within this context, a much greater number of solutions are found for
inhomogeneous waves than for homogeneous waves.
For instance, elliptical polarization is possible for homogeneous
waves only in two special directions 
(the `acoustic axes' \cite{BoHa92}), whereas it is possible in other 
directions for inhomogeneous waves.
Also, for certain inhomogeneous waves, the `complex scalar slowness',
which is the counterpart of the inverse of the speed for homogeneous
waves, may be arbitrarily prescribed.
In general, for a given orientation of the plane containing the 
directions of propagation and of attenuation, there is an infinity
of inhomogeneous wave solutions.

The paper is organized as follows.
First (section~\ref{basic_equations}) we recall the basic equations
describing the Mooney--Rivlin material. 
Then (section~\ref{Small_motions_superposed_on_large}) we write the 
equations governing the motion of a small-amplitude disturbance 
in a homogeneously deformed Mooney--Rivlin material. 
The incremental equations of motion and the incompressibility 
constraint are given. 
In order to solve these equations, we seek solutions of complex
exponential type, which are presented in section~\ref{vibrations}. 
The slowness bivector $\mathbf{S}$ is introduced; 
when the real and the imaginary parts of $\mathbf{S}$ are not
parallel, the wave is inhomogeneous, because the directions of
propagation and of attenuation do not coincide. 
The amplitude bivector $\mathbf{A}$ is also introduced; 
depending on whether or not the real and the imaginary parts of 
$\mathbf{A}$ are parallel, the wave is linearly or elliptically 
polarized, respectively. 
The special case of circular polarization corresponds to the 
`isotropy' of $\mathbf{A}$, that is $\mathbf{A \cdot A}=0$ 
\cite{Haye84}. 
Using these quantities, we then derive the incremental equations of 
motion for inhomogeneous plane waves of complex exponential type 
propagating in the deformed Mooney--Rivlin material.

Next (sections\ref{Cisotropic} and~\ref{Cnonisotropic})
the propagation of such disturbances is investigated.
Different sub-cases arise, depending on whether
or not the slowness complex vector $\mathbf{S}$ is isotropic,
and on whether the waves are circularly, elliptically,
or linearly polarized.
For each case, the general solution is provided
(amplitude and slowness bivectors,
incremental pressure, wave speed) as well as various explicit solutions
to the incremental equations of motion.

\section{Basic equations}
\label{basic_equations}

The Mooney--Rivlin material is a homogeneous isotropic hyperelastic
solid, for which the strain energy $ \Sigma$ per unit volume is given 
by \cite{Moon40}
\begin{equation}\label{Sigma}
2\Sigma=C(I-3) + D(\II-3),
\end{equation}
where  $C$, $D$ are material constants, and $I$, $\II$ are the first and
second invariants of the left Cauchy-Green strain tensor $\mathbb{B}$.
In order to satisfy the strong ellipticity condition, the constants
$C$, $D$ are such that \cite{Ogde70, BoHa95} either
$C>0, \, D \ge 0$ or $C \ge 0, \, D >0$.
When $D=0$, the material is said to be `neo-Hookean', but this
possibility is not considered in this paper.

The strain tensor $ \mathbb{B}$ is related to the deformation gradient
$ \mathbb{F}$ through $ \mathbb{B} = \mathbb{F} \mathbb{F}^\mathrm{T}$,
and its  first two invariants $I$ and $\II$ are defined by
$I=\rm{tr} \, \mathbb{B}$ and $2 \II=(\rm{tr}\mathbb{B})^2
-\rm{tr}(\mathbb{B}^2)$.

Because the Mooney--Rivlin material is incompressible, we have at
all times,
\begin{equation}\label{incompressibility}
  \mathrm{det} \, \mathbb{F}= (\mathrm{det} \, \mathbb{B})^{1/2}=1.
\end{equation}

Finally, the Mooney--Rivlin constitutive equation is derived from
\eqref{Sigma} and \eqref{incompressibility} as \cite{BoHa92}
\begin{equation} \label{Mooney-Rivlin}
\mbox{\boldmath $\sigma$}
 = -(p- D \II)\mathbf{1}+C \mathbb{B}-D \mathbb{B}^{-1},
\end{equation}
where \mbox{\boldmath $\sigma$} is the Cauchy stress and 
$-p \mathbf{1}$ a hydrostatic pressure, to be determined from the 
equations of motion, and initial and boundary conditions.

Throughout the paper, we assume that no body forces are applied,
so that the equations of motion reduce to
\begin{equation} \label{motion}
\text{div} \,  \mbox{\boldmath $\sigma$}
 =\rho(\partial ^2 \mathbf{x}/\partial t^2).
\end{equation}
Here, $ \mathbf{x}$ is the position at time $t$ of a point of the
deformed body which was at $ \mathbf{X}$ in the undeformed state, 
and $\rho$ is the constant mass density.

\section{Small motions superposed on a large homogeneous deformation}
\label{Small_motions_superposed_on_large}

In this Section we give the incremental equations of motion
corresponding to the superposition of a small-amplitude motion upon a
large homogeneous strain.
First, we assume that the Mooney--Rivlin material is subjected to a
finite pure homogeneous deformation, for which a point initially at 
$\mathbf{X}$ in the rectangular Cartesian coordinate system 
($O$, $\mathbf{i}$, $ \mathbf{j}$, $ \mathbf{k}$) moves to a point at 
$\mathbf{x}$ in the same system, with extension ratios
$\lambda_{\alpha}$ ($ \alpha =1,2,3$) and principal axes of deformation
along the axes of the  coordinate system, so that
\begin{equation}\label{stretch}
  x_{\alpha}=  \lambda_{\alpha} X_{\alpha} \quad
(\alpha =1,2,3; \text{ no sum}).
\end{equation}

The extension ratios are assumed to be distinct, constant, and 
without loss of generality, to be ordered
as $\lambda_{1} > \lambda_{2} > \lambda_{3}$.

The deformation gradient $\mathbb{F}$ and strain tensor $ \mathbb{B}$
corresponding to this finite static deformation are constant and given
by 
$\mathbb{F}= \mathrm{diag} \, (\lambda_1, \lambda_2, \lambda_3)$
and
$\mathbb{B}= \mathrm{diag} \, (\lambda_1^2, \lambda_2^2, \lambda_3^2)$.
Then the tensor $\mathbb{T}$, defined by \cite{Flav63, BoHa92}
\begin{equation} \label{T}
\begin{array}{l}
\mathbb{T}_{\alpha \beta} =0, \quad \alpha \ne \beta, \\
\mathbb{T}_{\alpha \alpha} = -p+ D(\lambda_1^{-2} + \lambda_2^{-2} +
\lambda_3^{-2}) +C \lambda_{\alpha}^{-2} -D \lambda_{\alpha}^{-2},
\quad (\alpha =1,2,3; \text{ no sum})
\end{array}
\end{equation}
where $p$ is  constant, is the constant Cauchy stress required to 
sustain the static deformation \eqref{stretch}.

Boulanger and Hayes \cite{BoHa92} have shown that two axes,
namely the `acoustic' axes  of the $\mathbb{B}^{-1}$-ellipsoid,
exhibit remarkable properties with respect to the propagation of 
finite-amplitude homogeneous plane waves in such a deformed 
Mooney--Rivlin material.
The acoustic axes are the only directions in which
circularly-polarized finite-amplitude homogeneous plane waves may 
propagate.
Their directions $\mathbf{n}^{\pm}$ are defined in terms of the
stretch ratios $\lambda_1$, $\lambda_2$,  $\lambda_3$, and of the 
unit vectors $\mathbf{i}$ and $\mathbf{k}$, which are in the 
directions of the principal axes of $\mathbb{B}$ (or 
$\mathbb{B}^{-1}$) corresponding to $\lambda_1$ and $\lambda_3$,
respectively. 
These directions are given by
\begin{equation}\label{n+-}
\mathbf{n^{\pm}}= \cos \phi \mathbf{i}  \pm \sin \phi \mathbf{k}, \,
\mathrm{where} \, \cos \phi = \sqrt{\frac
{\lambda_2^{-2}-\lambda_1^{-2}}
 {\lambda_3^{-2}-\lambda_1^{-2}}}, \,
\sin \phi =
\sqrt{\frac{\lambda_3^{-2}-\lambda_2^{-2}}
{\lambda_3^{-2}-\lambda_1^{-2}}},
\end{equation}
and are independent of the material parameters $C$ and $D$.
They lie along the normals to the planes of central circular sections
of the $\mathbb{B}^{-1}$-ellipsoid \cite[\S 5.7]{BoHa93}.
Their directions  $\mathbf{n}^{\pm}$ also appear in the Hamilton
cyclic decomposition of the $\mathbb{B}^{-1}$ tensor as,
\begin{equation} \label{hamil}
\mathbb{B}^{-1}= \lambda_2^{-2} \mathbf{1} - \textstyle{\frac{1}{2}}
(\lambda_3^{-2} - \lambda_1^{-2}) (\mathbf{n}^+ \otimes \mathbf{n}^-
+\mathbf{n}^- \otimes \mathbf{n}^+).
\end{equation} 

Now we consider a further deformation, possibly time-dependent, in
which the particle at $ \mathbf{x}$ moves to $\overline{\mathbf{x}}$
such that
\begin{equation} \label{xBar}
\overline{\mathbf{x}} = \mathbf{x}+\epsilon \mathbf{u}(\mathbf{x},t).
\end{equation}
Here the displacement $\epsilon \mathbf{u}$
 is a vector depending on $\mathbf{x}$ and on the time 
$t$, and $\epsilon$ is a small parameter.
Throughout the paper, we neglect terms of second and higher order 
in $\epsilon$.

Now if the deformation gradient $ \overline{\mathbb{F}}$, the strain
tensor  $ \overline{\mathbb{B}}$,  the invariants $\overline{I}$,
$\overline{\II}$, the pressure $ \overline{p}$, and the stress tensor
$\overline{\mathbb{T}}$ corresponding to the deformation \eqref{xBar}
are expanded in powers of $ \epsilon$ around their their value in the
static state of homogeneous deformation, we get
\begin{equation} \label{expansion}
\begin{array}{ll}
\overline{\mathbb{F}}= \mathbb{F} + \epsilon \mathbb{F}^* + \ldots,
\quad & \overline{\mathbb{B}}= \mathbb{B} + \epsilon \mathbb{B}^* +
\ldots , \\ \overline{I}= I + \epsilon I^* + \ldots, \quad &
\overline{\II}= \II + \epsilon \II^* + \ldots, \\
 \overline{p}= p + \epsilon p^* +  \ldots, \quad &
  \overline{\mathbb{T}}=\mathbb{T} + \epsilon \mathbb{T}^* + \ldots
\end{array}
\end{equation}
Then the equations of motion \eqref{motion}
written for $\overline{\mathbb{T}}$ and $\overline{\mathbf{x}}$ take
the form \cite{GrRS52}
\begin{equation} \label{mtn1}
\text{div} \, \mathbb{T}^* =\rho(\partial ^2
\mathbf{u}/\partial t^2),
\end{equation}
when we retain terms up to order $ \epsilon$. 
Explicitly, for the Mooney--Rivlin case, they are found to be (see also 
Belward \cite{Belw73}, Hayes and Horgan \cite{HaHo74} for the biaxial 
case):
\begin{equation} \label{mtn2}
\begin{array}{rr}
-p^*_{,1} +  C \lambda ^2_1 u_{1,11}+
              (C \lambda ^2_2 + D \lambda^{-2}_1)u_{1,22}+
              (C \lambda ^2_3 + D \lambda^{-2}_1)u_{1,33} & \\
 - D \lambda^{-2}_2 u_{2,12} - D \lambda^{-2}_3 u_{3,13}
         &  =   \rho \ddot{u}_1,             \vspace{3mm}     \\  
-p^*_{,2}   + (C \lambda ^2_1 + D \lambda^{-2}_2)u_{2,11}+
               C \lambda ^2_2 u_{2,22}+
              (C \lambda ^2_3 + D \lambda^{-2}_2)u_{2,33} & \\
- D \lambda^{-2}_1 u_{1,12} - D \lambda^{-2}_3 u_{3,23}
         &  =   \rho \ddot{u}_2,            \vspace{3mm}    \\ 
-p^*_{,3} + (C \lambda ^2_1 + D \lambda^{-2}_3)u_{3,11}+
              (C \lambda ^2_2 + D \lambda^{-2}_3)u_{3,22}+
              C \lambda ^2_3 u_{3,33} & \\
- D \lambda^{-2}_1 u_{1,13} - D \lambda^{-2}_2 u_{2,23}
         &  =   \rho \ddot{u}_3,
\end{array}
\end{equation}
where commas and dots denote differentiation with respect to position 
$\mathbf{x}$ and time $t$, respectively.
Thus, for example, $u_{1,2}= \partial u_1 / \partial y$.

Finally, the incompressibility constraint \eqref{incompressibility}
yields \cite{GrRS52}
 \begin{equation} \label{div_u}
u_{1,1} + u_{2,2} + u_{3,3} = 0.
\end{equation}

\section{Vibrations of complex exponential type}
\label{vibrations}

In order to solve the equations governing the incremental motion,
we assume that the displacement $\epsilon \mathbf{u}$
and the incremental pressure $\epsilon p^*$ are of the form
\begin{equation} \label{expS}
\mathbf{u}  =  \textstyle{\frac{1}{2}}
  \{ \mathbf{A}e^{i\omega(\mathbf{S \cdot x}-t)} + \mathrm{c.c.} \},
\quad
p^* =   \textstyle{\frac{1}{2}}
\{ i\omega Q e^{i\omega (\mathbf{S \cdot x}-t)} + \mathrm{c.c.}  \},
\end{equation}
where $\mathbf{A}=\mathbf{A'}+ i \mathbf{A''}$ is a complex vector,
called the `amplitude bivector' \cite{BoHa93}, $\omega$ is the real
frequency, $Q$ is a scalar, $\mathbf{S} = \mathbf{S'} +i \mathbf{S''}$
is the `slowness bivector', and `c.c.' stands for `complex conjugate'.

An ellipse may be associated with a bivector, as the ellipse for
which the vector corresponding to the real part and the 
vector corresponding to the imaginary part are conjugate vectors.
That is, to a bivector $\mathbf{D}=\mathbf{D'}+ i \mathbf{D''}$ say,
we associate the ellipse described by the point $M$ such that 
$\mathbf{OM} =  \mathbf{D'}\cos \theta +  \mathbf{D''} \sin \theta$, 
$0 \le \theta \le 2 \pi$.

When the real and imaginary parts of  
$\mathbf{A}$ are parallel, that is, when the ellipse of $\mathbf{A}$
degenerates into a segment,
the wave is linearly-polarized along their common direction.
Otherwise, the wave is elliptically-polarized, and the ellipse of 
polarization is the ellipse of $\mathbf{A}$.
The special case of circular polarization corresponds to the 
`isotropy' of $\mathbf{A}$ \cite{Haye84}
\begin{equation}\label{circular}
\mathbf{A \cdot A}=0.
\end{equation}

The planes defined by $\mathbf{S' \cdot x}=$ constant are called
planes of constant phase, and the planes defined by 
$\mathbf{S'' \cdot x}=$ constant are the planes of constant amplitude.
When the real part $\mathbf{S'}$  and the imaginary part 
$\mathbf{S''}$ of $\mathbf{S}$ are not parallel,
the wave is said to be inhomogeneous.
In that case, we introduce the `directional ellipse' of the slowness 
bivector $\mathbf{S}$, defined as follows.
Let $\mathbf{\widehat{m}}$ and $\mathbf{\widehat{n}}$ be
unit vectors along the respective major and minor semi-axes of
the ellipse of $\mathbf{S}$, and $m$ be the aspect ratio of the 
ellipse of $\mathbf{S}$.
Then the directional ellipse of the slowness bivector $\mathbf{S}$ 
is the ellipse of the `propagation bivector' $\mathbf{C}$, defined by
\begin{equation} \label{C}
\mathbf{C}=m\mathbf{\widehat{m}}+i\mathbf{\widehat{n}},
\quad \mathrm{with} \quad
\mathbf{\widehat{m} \cdot \widehat{m}}
=\mathbf{\widehat{n} \cdot \widehat{n}}=1 ,
\quad \mathbf{\widehat{m} \cdot \widehat{n}}=0.
\end{equation}
Then $\mathbf{S}$ may be written as \cite{Haye84}
$\mathbf{S}=N\mathbf{C}
=N(m\mathbf{\widehat{m}}+i\mathbf{\widehat{n}})$,
where $N= N' + i N''$ is a complex number, called the `complex scalar
slowness'.
Then the directions of propagation and attenuation are the directions 
of $ \mathbf{S'}$ and $\mathbf{S'}$, respectively, given by
\begin{equation}\label{S'S''}
\mathbf{S'}= m N'\mathbf{\widehat{m}} -  N''\mathbf{\widehat{n}},
 \quad
\mathbf{S''}= m N''\mathbf{\widehat{m}} +  N'\mathbf{\widehat{n}}.
\end{equation}
Note that
\begin{equation}\label{S'.S''}
\mathbf{S'  \cdot  S''}= (m^2 -1) N' N'',
\end{equation}
and therefore, the planes of constant phase are orthogonal to the
planes of constant amplitude either when $ m=1$ or when $N$ is purely
real or purely imaginary.

So, we may write solutions of the form \eqref{expS} in terms of 
$\mathbf{C}$ as
\begin{equation} \label{expC}
\mathbf{u} =   \textstyle{\frac{1}{2}}
\{ \mathbf{A}e^{i\omega  (N\mathbf{C \cdot x}-t)} +\mathrm{c.c.}\},
\quad
p^* =  \textstyle{\frac{1}{2}}
\{ i\omega N P e^{i\omega  (N\mathbf{C \cdot x}-t)} +\mathrm{c.c.}\},
\end{equation}
where $P=N^{-1}Q$. 
Now we substitute these expressions into \eqref{mtn2} and~\eqref{div_u}.

First, equation \eqref{div_u} imposes the condition
\begin{equation} \label{A.C}
\mathbf{A \cdot C} =0.
\end{equation}
A geometrical interpretation  \cite[\S 2.4]{BoHa93} of this equation
is that the orthogonal projection of the ellipse of the bivector
$\mathbf{C}$ onto the plane of the bivector $\mathbf{A}$ is similar 
and similarly situated to the ellipse of $\mathbf{A}$, rotated 
through a quadrant.
The amplitude and propagation bivectors are said to be `orthogonal'.
For homogeneous plane waves, the equation reduces to 
$\mathbf{A \cdot n}=0$ (where $\mathbf{n}$ is a real vector in the 
direction of propagation), which simply means that the polarization 
of the wave is transverse.

Next, the equations of motion \eqref{mtn2} yield
\begin{equation} \label{mtn3}
-P \mathbf{C} 
 + C (\mathbf{C \cdot } \mathbb{B} \mathbf{C}) \mathbf{A} 
 + D[(\mathbf{C \cdot C})\mathbb{B}^{-1} \mathbf{A} 
	-(\mathbf{A \cdot } \mathbb{B}^{-1} \mathbf{C})\mathbf{C}] 
=\rho N^{-2}\mathbf{A}.
\end{equation}
Taking the dot product of this last equation with $\mathbf{C}$,
and using equation \eqref{A.C}, yields
\begin{equation} \label{mtn4}
-P (\mathbf{C \cdot C})=0.
\end{equation}

In conclusion, the propagation of small-amplitude waves of complex
exponential type in a homogeneously deformed Mooney--Rivlin material is
governed by the following equations,
\begin{equation} \label{mtn0}
\left\{
\begin{array}{l}
-P \mathbf{C} + C (\mathbf{C \cdot } \mathbb{B} \mathbf{C}) \mathbf{A}
+ D [(\mathbf{C \cdot C}) \mathbf{1} - \mathbf{C} \otimes \mathbf{C}]
\mathbb{B}^{-1} \mathbf{A}
=\rho N^{-2}\mathbf{A}, \\
\mathbf{A \cdot C}=0, \quad
P (\mathbf{C \cdot C})=0.
\end{array}
\right.
\end{equation}

Now, we treat in turn the case where $\mathbf{C}$ is isotropic
($\mathbf{C \cdot C}=0$) and the  case where $\mathbf{C}$ is not 
isotropic ($\mathbf{C \cdot C} \ne 0$).

\section{Propagating evanescent solutions, $\mathbf{C \cdot C}=0$}
\label{Cisotropic}

Here it is seen that corresponding to \textit{any} isotropic
bivector $\mathbf{C}$, there exists an infinity of linearly,
circularly, and elliptically polarized inhomogeneous plane wave 
solutions.

First, we note that when $\mathbf{C \cdot C}=0$, we have $m=1$ in 
\eqref{C} and  we deduce from \eqref{S'.S''} that the planes of 
constant phase are orthogonal to the planes of constant amplitude.

Next, the equations \eqref{mtn3} and \eqref{mtn4} reduce to
\begin{equation} \label{mtniso}
\begin{array}{l}
-P \mathbf{C} 
 + C (\mathbf{C \cdot } \mathbb{B} \mathbf{C}) \mathbf{A} 
  - D (\mathbf{A \cdot } \mathbb{B}^{-1} \mathbf{C})\mathbf{C}
=\rho N^{-2}\mathbf{A} ,  \\ 
\mathbf{A \cdot C}= 0, \quad \mathbf{C \cdot C} = 0.
\end{array}
\end{equation}

Equations $\eqref{mtniso}_{2,3}$ allow us to decompose $\mathbf{A}$ as
 \cite[\S 2.9]{BoHa93}
\begin{equation} \label{decompiso}
\mathbf{A}= \alpha_1 \mathbf{C} +\alpha_2 \mathbf{C} \wedge
\widetilde{\mathbf{C}},
\end{equation}
where $\alpha_{1}$, $\alpha_{2}$ are real constants and
$\widetilde{\mathbf{C}}$ is the complex conjugate of $\mathbf{C}$.
Note that because $\mathbf{C}$ is isotropic, it is written as 
$\mathbf{C} = \mathbf{\hat{m}} + i \mathbf{\hat{n}}$ 
(\eqref{C} with $m=1$) and so,
$\mathbf{C} \wedge \widetilde{\mathbf{C}}= 2i \mathbf{\hat{n}} \wedge
\mathbf{\hat{m}}$ is parallel to a real vector.
Substituting \eqref{decompiso} in $\eqref{mtniso}_1$ leads to two 
different types of solutions.

\subsubsection*{(i)  First type of solution:
    linearly and elliptically polarized waves}

Here the amplitude bivector $\mathbf{A}$ is given by
$\mathbf{A}= \alpha_1 \mathbf{C} +\alpha_2 \mathbf{C} \wedge
\widetilde{\mathbf{C}}$, $\alpha_2 \neq 0$,
where $\alpha_{1}$, $\alpha_{2}$ are real constants.
The corresponding complex scalar slowness $N$ is given by
\begin{equation} \label{CBC}
\rho N^{-2} = C (\mathbf{C \cdot }\mathbb{B}\mathbf{C}),
\end{equation}
and the incremental pressure is given by \eqref{expC}, where
\begin{equation}
P = - \alpha_1 D (\mathbf{C \cdot } \mathbb{B}^{-1} \mathbf{C}) 
- \alpha_2 D (\mathbf{C} \wedge \widetilde{\mathbf{C}}) 
 \cdot  \mathbb{B}^{-1} \mathbf{C}.
\end{equation}
Note that in this case, $\mathbf{A \cdot A}= -\alpha_2^2
(\mathbf{C} \cdot \widetilde{\mathbf{C}})^2 \ne 0$, and thus the wave 
is not circularly-polarized. 
When $ \alpha_1=0$, the wave is linearly-polarized in the direction 
of $ \mathbf{C} \wedge \widetilde{\mathbf{C}}$.
When $\alpha_1 \ne 0$, the wave is elliptically-polarized.
Also, note that \eqref{CBC} can be written in terms of the slowness
bivector $\mathbf{S}$ as 
$\rho =C \mathbf{S \cdot } \mathbf{B}\mathbf{S}$.
The imaginary part of this equation yields
$ \mathbf{S'}  \cdot  \mathbb{B} \mathbf{S''}=0$,
which means that the direction of the normal to the planes of equal 
phase and the direction of the normal to the planes of equal amplitude
are conjugate directions with respect to the $ \mathbb{B}$-ellipsoid.
This condition has been established previously for linearly-polarized 
finite-amplitude inhomogeneous plane waves propagating in a deformed 
Mooney--Rivlin material with an isotropic slowness bivector 
\cite{Dest00}.

\subsubsection*{(ii) Second type of solution: circularly-polarized waves}

Here the amplitude bivector $\mathbf{A}$ is given by
$\mathbf{A}= \alpha_1 \mathbf{C}$, $\alpha_1 \neq 0$,
where $\alpha_{1}$ is a real constant, and the incremental pressure is
given by \eqref{expC}, where
\begin{equation}
P = \alpha_1 [C (\mathbf{C \cdot } \mathbb{B} \mathbf{C}) -
D (\mathbf{C \cdot } \mathbb{B}^{-1} \mathbf{C}) - \rho N^{-2}].
\end{equation}
For this wave, the  complex scalar slowness $N$ is
\textit{arbitrary} (a similar situation was encountered by 
Boulanger and Hayes \cite{BoHa96});
thus the displacement, which is the real part of
$\epsilon 
  \{ \alpha_1 \mathbf{C} \exp i\omega (N \mathbf{C \cdot x} -t) \}$
is independent of the material constants $C$ and $D$.
Also, $\mathbf{A \cdot A}=\alpha_1^2 (\mathbf{C \cdot C}) = 0$
and the wave is circularly-polarized.

Because there is an infinity of choices for an isotropic bivector
$\mathbf{C}$, there is a triple infinity of propagation, attenuation,
and polarization directions for linearly (type (i), $\alpha_1 =0$),
elliptically (type (i), $\alpha_1 \ne 0$), and circularly (type (ii))
polarized inhomogeneous plane waves of complex exponential type,
provided the slowness bivector is isotropic.
This is in sharp contrast to  elliptically and circularly polarized 
\textit{homogeneous} plane waves, which can only propagate along an
acoustic axis \cite{BoHa92}.

Now we present explicit examples of the two types of solution 
corresponding to an isotropic propagation
bivector $\mathbf{C}$.

\subsubsection*{Example 1: Waves with an isotropic slowness bivector}

We choose an isotropic bivector $\mathbf{C}$ and write the
corresponding two types of solutions.
One set of solutions consists of elliptically-polarized waves,
the other set consists of circularly-polarized waves travelling with
an arbitrary speed.
Let $\mathbf{C}=(3 \mathbf{i} +5 i \mathbf{j} +4\mathbf{k})/5$.

For waves of type (i), we have, using \eqref{decompiso}, 
$\mathbf{A} =
(3\alpha_1 + 4i \alpha_2) \mathbf{i} + 5 i \alpha_1 \mathbf{j}
+(4\alpha_1 - 3i \alpha_2) \mathbf{k}$, 
$\rho N^{-2}= 
C(9\lambda_1^2 -25\lambda_2^2+16\lambda_3^2)/25$ 
and 
$P= 
- D [(9\lambda_1^2 -25\lambda_2^2+16\lambda_3^2)\alpha_1
      + 12i ( \lambda_1^{-2} -\lambda_3^{-2}) \alpha_2]/5$. 

If the primary finite deformation is such that 
$9\lambda_1^2 -25\lambda_2^2 +16\lambda_3^2 > 0$, then $N$ is 
real and we can write the following solution to the incremental 
equations of motion in a deformed Mooney--Rivlin material
\eqref{mtn2}:
\begin{equation}
\begin{array}{l} \label{IsoEx1a}
u_1 = e^{- \omega |N| y}
\{3\alpha_1  \cos\omega  [|N| (3x+4z)/5-  t]
-4\alpha_2 \sin\omega [|N|(3x+4z)/5-t]\},
\vspace{3pt}\\
u_2 = -5\alpha_1 e^{- \omega  |N| y}
\sin \omega [|N|(3x+4z)/5 - t], \vspace{3pt}\\
u_3 = e^{- \omega |N| y}
\{4\alpha_1  \cos\omega [  |N|(  3x+4z)/5-  t]
+3\alpha_2 \sin\omega [|N|(3x+4z)/5-t]\}, \vspace{3pt}\\
\begin{split}
p^* =\fifth D \omega |N| e^{- \omega |N| y} 
\{ \alpha_1 (9\lambda_1^{-2} - & 25 \lambda_2^{-2}+16\lambda_3^{-2})
   \sin\omega [|N|(3x+4z)/5-t]  \\
 + & 12 \alpha_2(\lambda_1^{-2}
-\lambda_3^{-2}) \cos\omega [|N|(3x+4z)/5-t]\},
\end{split}
\end{array}
\end{equation}
where $\omega, \alpha_1, \alpha_2$ are arbitrary ($\alpha_2 \neq 0$), 
and 
\begin{equation}
|N|= 5\sqrt{\rho / [C (9\lambda_1^2 -25\lambda_2^2
+16\lambda_3^2)]}. \label{IsoExN}
\end{equation}
This wave travels in the direction of $3 \mathbf{i} + 4 \mathbf{k}$
and is attenuated in the $y$-direction.
If $9\lambda_1^2 -25\lambda_2^2 +16\lambda_3^2 > 0$, then a solution 
is given by
\begin{equation}
\begin{array}{l} \label{IsoEx1b}
u_1 = e^{- \omega |N| (3x+4z)/5}
\{3\alpha_1  \cos\omega  [- |N| y -  t]-4\alpha_2 \sin\omega [-|N|y -t]\},
\vspace{3pt}\\
u_2 = -5\alpha_1 e^{- \omega  |N| (3x+4z)/5}
\sin \omega [- |N|y - t], \vspace{3pt}\\
u_3 = e^{- \omega |N| (3x+4z)/5}
\{4\alpha_1  \cos\omega [-|N|y -  t]
+3\alpha_2 \sin\omega [-|N|y - t]\}, \vspace{3pt}\\
\begin{split}
p^* = \fifth D \omega |N| e^{- \omega |N| (3x+4z)/5} 
\{ \alpha_1 (9\lambda_1^{-2} & -25\lambda_2^{-2}+16\lambda_3^{-2})
   \sin\omega [-|N|y - t]  \\
 + & 12 \alpha_2(\lambda_1^{-2}
-\lambda_3^{-2}) \cos\omega [-|N|y - t]\},
\end{split}
\end{array}
\end{equation}
where $\omega, \alpha_1, \alpha_2$ are arbitrary ($\alpha_2 \neq 0$), 
and $|N|$ is again given by \eqref{IsoExN}.
This wave travels in the direction of $- \mathbf{j}$ and is attenuated
in the direction of $3 \mathbf{i} + 4 \mathbf{k}$.
In both cases \eqref{IsoEx1a} and \eqref{IsoEx1b}, the wave propagates
with speed $|N|^{-1}$ where $|N|$ is given by \eqref{IsoExN}, 
and is elliptically-polarized (two conjugate radii of the ellipse are 
$\alpha_1(3\mathbf{i}+4\mathbf{k})$ and $ 5 \alpha_1 \mathbf{j}
+\alpha_2(4\mathbf{i} -3 \mathbf{k})$).

For waves of type (ii), we have: 
$\mathbf{A} = 
 \alpha_1(3\mathbf{i} +5i\mathbf{j} +4\mathbf{k})$, 
$\rho N^{-2}$ is arbitrary and 
$P=
 \alpha_1  [C (9\lambda_1^2 - 25\lambda_2^2+16\lambda_3^2) 
   - D (9\lambda_1^{-2} -25 \lambda_2^{-2}+16\lambda_2^{-2})
      - \rho N^{-2}]/5$.
For $N$ real, the corresponding solution is given by
\begin{equation} \label{IsoEx2}
\begin{array}{l}
u_1 = 3\alpha_1 e^{- \omega N y} \cos\omega [N(3x+4z)/5-t], \\
u_2 = -5\alpha_1 e^{-\omega N y}  \sin \omega  [N(3x+4z)/5-t],\\
u_3 = 4\alpha_1 e^{- \omega N y} \cos \omega [N(3x+4z)/5-t],\\
\begin{split}
p^* = \fifth\alpha_1\omega N e^{- \omega N y} 
     \{ \rho
& N^{-2} - C(9\lambda_1^2-25\lambda_2^2+16 \lambda_3^2)  \\ 
+ & D(9\lambda_1^{-2}-25\lambda_2^{-2}+16\lambda_3^{-2}) \}
   \sin \omega  [N (3x+4z)/5 -t],
\end{split}
\end{array}
\end{equation}
where $\alpha_1, \omega$ and $N$ are arbitrary. 
This wave travels in the direction of $3 \mathbf{i} + 4 \mathbf{k}$ 
with an arbitrary speed $N^{-1}$, is attenuated in the $y$-direction, 
and is circularly-polarized (two orthogonal radii of the circle are 
$\alpha_1(3\mathbf{i}+4\mathbf{k})$ and $ 5 \alpha_1 \mathbf{j}$). 
The displacement field $ \epsilon \mathbf{u}$ does not depend on the
constants $C$ and $D$. 
It is easily checked that the solutions \eqref{IsoEx1a}, 
\eqref{IsoEx1b}, and \eqref{IsoEx2} satisfy the general equations of 
motion \eqref{mtn2}.


\section{Propagating evanescent solutions, $\mathbf{C \cdot C} \ne 0$}
\label{Cnonisotropic}

In this Section, we consider the case where the bivector $\mathbf{C}$
is not isotropic.
A great variety of solutions is uncovered, and a systematic method 
of construction and classification for linearly, circularly, and
elliptically polarized waves is presented. 

When $\mathbf{C \cdot C} \neq 0$, \eqref{mtn0}$_3$ implies 
that $P=0$ and therefore $p^*=0$.
Equation \eqref{mtn0}$_1$ then reduces to
\begin{equation} \label{mtnP=0}
\{C (\mathbf{C \cdot } \mathbb{B} \mathbf{C}) \mathbf{1}
+D [(\mathbf{C \cdot C}) \mathbf{1}-
\mathbf{C} \otimes \mathbf{C}] \mathbb{B}^{-1} \}
\mathbf{A} =\rho N^{-2}\mathbf{A}.
\end{equation}
Following \cite{BoHa92}, we symmetrize this
equation, using \eqref{A.C}, to give the equivalent form
\begin{equation} \label{mtnNiso}
\mbox{\boldmath $\Pi$}
 [C (\mathbf{C \cdot } \mathbb{B} \mathbf{C}) \mathbf{1} 
	+ D (\mathbf{C \cdot C}) \mathbb{B}^{-1}] 
\mbox{\boldmath $\Pi$} \mathbf{A} =\rho N^{-2}\mathbf{A}
\quad\mbox{and}\quad 
\mathbf{A \cdot C} = 0,
\end{equation}
where we have introduced the `complex' projection operator 
$\mbox{\boldmath $\Pi$}$, defined by
\begin{equation}\label{Pi}
\mbox{\boldmath $\Pi$}
= \mathbf{1} 
- \frac{\mathbf{C} \otimes \mathbf{C}}{\mathbf{C}  \cdot \mathbf{C}}.
\end{equation}
This operator generalizes the `real' projection 
$\mathbf{1}-\mathbf{n}\otimes \mathbf{n}$ upon the plane 
$\mathbf{n \cdot x}=0$ \cite{BoHa92}, and has the following properties:
$\mbox{\boldmath $\Pi$}^2= \mbox{\boldmath $\Pi$}$, 
$\mbox{\boldmath $\Pi$} \mathbf{C}=0$ and 
$\mbox{\boldmath $\Pi$} \mathbf{A} =\mathbf{A}$.

By inspection of \eqref{mtnNiso}, we see that solving the
equations of motion, once $\mathbf{C}$ is prescribed, is equivalent to
finding the eigenbivectors $\mathbf{A}$ of the tensor 
$\mbox{\boldmath $\Pi$}[C(\mathbf{C \cdot } \mathbb{B} \mathbf{C}) 
  \mathbf{1} + D (\mathbf{C \cdot C}) \mathbb{B}^{-1}] 
    \mbox{\boldmath $\Pi$}$ such that $\mathbf{A \cdot C} = 0$, 
and their corresponding eigenvalues $\rho N^{-2}$. 
This procedure is analogous to that used in 
\cite{BoHa92} for finite-amplitude homogeneous plane waves, with the 
replacement of their real vectors of propagation $\mathbf{n}$ and 
polarization $\mathbf{a}$, and their speed $c$ with the bivectors 
$\mathbf{C}$ and $\mathbf{A}$, and the quantity $N^{-1}$, respectively.

First of all, we compute the possible eigenvalues $\rho N^{-2}$.

\subsection{Secular equation}
\label{Secular_equation}

The equations \eqref{mtnNiso} admit solutions, provided that
\begin{equation} \label{detM=0}
\mathrm{det} (\mbox{\boldmath $\Pi$}
 [C (\mathbf{C \cdot } \mathbb{B} \mathbf{C}) \mathbf{1}
    + D (\mathbf{C \cdot C}) \mathbb{B}^{-1}] \mbox{\boldmath $\Pi$}
- \rho N^{-2}\mathbf{1})=0.
\end{equation}
Equation \eqref{detM=0} is the classical secular equation for
inhomogeneous plane waves. 
Because $\mathrm{det} \, \mbox{\boldmath $\Pi$} =0$, then 
$\rho N^{-2}=0$ is one root of \eqref{detM=0}. 
The two other eigenvalues $\rho N_{\pm}^{-2}$ (say) are the roots of 
the quadratic
\begin{eqnarray}
[ \rho N^{-2}- C (\mathbf{C \cdot } \mathbb{B} \mathbf{C})]^2 
- D[(\mathbf{C \cdot C}) \mathrm{tr} \, 
      \mathbb{B}^{-1} -(\mathbf{C \cdot }\mathbb{B}^{-1} \mathbf{C})] 
[\rho N^{-2}- C (\mathbf{C \cdot } \mathbb{B} \mathbf{C})] 
&&\nonumber\\
\mbox{}+D^2 (\mathbf{C \cdot C})(\mathbf{C \cdot } \mathbb{B}
\mathbf{C}) &=&0.
 \label{secular}
\end{eqnarray}
Explicitly,  $\rho N_{\pm}^{-2}$ are given in terms of $\mathbf{C}$
alone as
\begin{multline} \label{rho+-}
2 \rho N^{-2}_{\pm}= 2 C (\mathbf{C \cdot } \mathbb{B} \mathbf{C}) 
+ D [(\mathbf{C \cdot C}) \mathrm{tr} \, \mathbb{B}^{-1}
 -(\mathbf{C \cdot } \mathbb{B}^{-1} \mathbf{C})] \\
 \pm D \sqrt{[(\mathbf{C \cdot C}) \mathrm{tr} \, \mathbb{B}^{-1}
-(\mathbf{C \cdot } \mathbb{B}^{-1} \mathbf{C})]^2 
- 4(\mathbf{C \cdot C})(\mathbf{C \cdot } \mathbb{B}\mathbf{C})}.
\end{multline}
These last quantities are the two eigenvalues corresponding to the
amplitude bivectors which are orthogonal to $\mathbf{C}$.
Equations \eqref{rho+-} generalize the corresponding equations
written for homogeneous waves by Boulanger and Hayes \cite{BoHa95},
by replacing their vector of propagation $\mathbf{n}$ and speed $c$ by
$\mathbf{C}$ and $N^{-1}$, respectively.
Now we establish in turn the conditions for linear, circular,
and elliptical polarization.

\subsection{Linearly-polarized waves, $\mathbf{C \cdot C} \ne 0$}
\label{Linearly_polarized_waves_Cnonisotropic}

Here we prove that linearly-polarized inhomogeneous waves with a
nonisotropic slowness bivector can propagate in a deformed
Mooney--Rivlin material only when they are polarized in a principal
direction.

When the amplitude bivector $\mathbf{A}$ is parallel to a real unit
vector $\mathbf{a}$ (linear polarization), the incompressibility 
constraint \eqref{A.C} written as $ \mathbf{a \cdot S}=0$ implies 
that the directions of propagation ($\mathbf{S'}$) and attenuation 
($\mathbf{S''}$) are both orthogonal to $\mathbf{a}$. 
Hence, we write $\mathbf{a}=\alpha \mathbf{S'} \wedge \mathbf{S''}$, 
where $\alpha$ is such that $ \mathbf{a \cdot a}=1$. 
The equations of motion \eqref{mtn0} then reduce to
\begin{equation} \label{linear}
C ( \mathbf{S \cdot } \mathbb{B} \mathbf{S}) \mathbf{a} 
	+ D[( \mathbf{S \cdot S})
\mathbf{1} - \mathbf{S} \otimes \mathbf{S}] \mathbb{B}^{-1}\mathbf{a}
 = \rho \mathbf{a}, \quad
 \mathbf{a}=\alpha \mathbf{S'} \wedge \mathbf{S''}, \quad
 P=0.
\end{equation}
The dot product of \eqref{linear}$_1$ by 
$\mathbf{\widetilde{S}} =\mathbf{S'} -i \mathbf{S''}$ yields
\begin{equation}
(\mathbf{S \cdot S}) 
 (\mathbf{a \cdot }\mathbb{B}^{-1} \mathbf{\widetilde{S}})
- (\mathbf{S \cdot \widetilde{S}}) 
    (\mathbf{a \cdot }\mathbb{B}^{-1} \mathbf{S})=0,
\end{equation}
or, separating real and imaginary parts,
\begin{equation}
\left\{
\begin{array}{l}
(\mathbf{S''  \cdot S''}) 
 ( \mathbf{a \cdot } \mathbb{B}^{-1} \mathbf{S'})
- (\mathbf{S' \cdot  S''}) 
   ( \mathbf{a \cdot }\mathbb{B}^{-1} \mathbf{S''})=0,\\
(\mathbf{S'  \cdot S''}) 
 (\mathbf{a \cdot }\mathbb{B}^{-1} \mathbf{S'})
   -(\mathbf{S' \cdot  S'}) 
       ( \mathbf{a \cdot }\mathbb{B}^{-1} \mathbf{S''})=0.
\end{array}
\right.
\end{equation}
Using $(\mathbf{S' \cdot S'})(\mathbf{S'' \cdot S''})
 -(\mathbf{S'  \cdot S''})^2 = \alpha^{-2} \mathbf{a \cdot a} \ne 0$, 
we see that this homogeneous linear system of two equations admits 
only trivial solutions, that is
$\mathbf{a \cdot }\mathbb{B}^{-1} \mathbf{S'}
 = \mathbf{a \cdot }\mathbb{B}^{-1} \mathbf{S''}=0$.
Thus $\mathbf{a}$ is parallel to $\mathbb{B}^{-1} \mathbf{S'} \wedge
\mathbb{B}^{-1} \mathbf{S''}$, so that for some $ \beta$,
\begin{equation}
\mathbf{a}=\beta (\mathbb{B}^{-1} \mathbf{S'} \wedge \mathbb{B}^{-1}
\mathbf{S''})=  \beta \mathbb{B} (\mathbf{S'} \wedge
\mathbf{S''})=\alpha^{-1}  \beta \mathbb{B} \mathbf{a}.
\end{equation}
This last equality means that $\mathbf{a}$ is an eigenvector of~$ \mathbb{B}$.
Therefore, provided $\mathbf{S}$ is not 
isotropic, linearly-polarized inhomogeneous plane waves of complex 
exponential type can propagate in a deformed Mooney--Rivlin material 
only when polarized in a principal direction of the primary 
homogeneous deformation.
Examples of such waves are given in
section~\ref{Elliptically_polarized_waves_Cnonisotropic}.
When $\mathbf{S}$ (or equivalently, $\mathbf{C}$) is isotropic,
the inhomogeneous plane wave can be linearly-polarized in
nonprincipal directions (such as in example \eqref{IsoEx1a}
when $ \alpha_1 =0$).

\subsection{Circularly-polarized waves, $\mathbf{C \cdot C} \ne 0$}
\label{Circularly_polarized_waves_Cnonisotropic}

Here, we establish the condition for the propagation of
circularly-polarized inhomogeneous plane waves with a nonisotropic
slowness bivector.
  From \eqref{circular} and \eqref{mtnP=0}, the equations governing the
propagation of such waves are
\begin{equation} \label{circNiso}
\begin{array}{l}
C (\mathbf{C \cdot } \mathbb{B} \mathbf{C}) \mathbf{A}
 +D (\mathbf{C \cdot C})
\mathbb{B}^{-1}  \mathbf{A}- D (\mathbf{A} \cdot   \mathbb{B}^{-1}
\mathbf{C})  \mathbf{C} =\rho N^{-2}\mathbf{A}, \\
\mathbf{A \cdot C} =0, \quad  \mathbf{A \cdot A} = 0.
\end{array}
\end{equation}
Taking the dot product of \eqref{circNiso}$_1$ by $\mathbf{A}$ yields
$\mathbf{A \cdot }\mathbb{B}^{-1}\mathbf{A}=0$, or, using~\eqref{hamil},
\begin{equation} \label{A.n+-=0}
\mathbf{A \cdot } \mathbf{n}^{\pm}=0.
\end{equation}
Therefore, $\mathbf{A}$ is orthogonal to $ \mathbf{n}^{\pm}$.
In other words, the planes of circular polarization for
inhomogeneous plane waves with a nonisotropic slowness bivector are
the planes of central circular section of the
$\mathbb{B}^{-1}$-ellipsoid.
This is the same situation as for homogeneous plane waves~\cite{BoHa92}.

  From \eqref{circNiso}$_{2,3}$ and \eqref{A.n+-=0} it follows that the
projection of the ellipse of $\mathbf{C}$ onto a plane of central
circular section of the $\mathbb{B}^{-1}$-ellipsoid is a circle.
Note that there is an infinity of such $\mathbf{C}$.

We consider the case where $ \mathbf{A \cdot } \mathbf{n}^+=0$ and 
denote the corresponding isotropic amplitude bivector  by
$\mathbf{A}^{\theta}_\odot$.
Noting that $\mathbf{n}^+$ is defined by \eqref{n+-}, 
it follows that $\mathbf{A}^{\theta}_\odot$ is given 
by \cite[\S 2.2]{BoHa93}
\begin{equation} \label{Adot}
\mathbf{A}^{\theta}_\odot= \alpha e^{i\theta}(\mathbf{j} \wedge
\mathbf{n^+} + i\mathbf{j}), \quad \mbox{$\alpha$, $\theta$ real}.
\end{equation}

Now, because $\mathbf{A}^{\theta}_\odot$,
$\widetilde{\mathbf{A}}^{\theta}_\odot$, and $ \mathbf{n^+}$ are
linearly independent bivectors, the corresponding propagation bivector
$\mathbf{C}^{\theta}_\odot$ (say) can be written as
$\mathbf{C}^{\theta}_\odot= \lambda \mathbf{A}^{\theta}_\odot + \mu
\widetilde{\mathbf{A}}^{\theta}_\odot + \nu \mathbf{n^+}$.
  From \eqref{circNiso}$_2$ we have $\mathbf{C}^{\theta}_\odot  \cdot 
\mathbf{A}^{\theta}_\odot = \mu \widetilde{\mathbf{A}}^{\theta}_\odot
 \cdot \mathbf{A}^{\theta}_\odot =0$, and therefore $\mu =0$.
Then from \eqref{C}, we have $\mathbf{C}^{\theta}_\odot .
\mathbf{C}^{\theta}_\odot = m^2 -1 =\nu^2$, and therefore, $\nu$ is
real and given by $ \nu = \sqrt{m^2-1}$. 
Also,
$\mathbf{C}^{\theta}_\odot \cdot 
    \widetilde{\mathbf{C}}^{\theta}_\odot =
      m^2 +1 = 2 \lambda \widetilde{\lambda} \alpha^2 + \nu^2$, 
and so $|\lambda \alpha | =1$.
We conclude that $\mathbf{C}$ can be written as
\begin{equation}  \label{Cdot}
\mathbf{C}^{\theta}_\odot = e^{i\theta}(\mathbf{j} \wedge \mathbf{n^+}
+ i\mathbf{j}) + \sqrt{m^2 -1} \mathbf{n^+}.
\end{equation}

Using \eqref{Adot} and  \eqref{Cdot}, we compute the eigenvalue 
$\rho (N^{\theta}_\odot)^{-2}$ (say) from  \eqref{circNiso}$_1$ and 
find that it is given by
\begin{equation} \label{Ndot}
\rho(N^{\theta}_\odot)^{-2}= \Lambda^{\theta}_\odot 
[ C \Lambda^{\theta}_\odot + D \sqrt{m^2 -1} ],
\end{equation}
where
\begin{equation}  \label{Lambdadot}
\Lambda^{\theta}_\odot =  \lambda_2^{-2}  \sqrt{m^2 -1} +
e^{i\theta} \sqrt{
(\lambda_2^{-2}-\lambda_1^{-2}) (\lambda_3^{-2}-\lambda_2^{-2})}.
\end{equation}

\subsubsection*{Example 2: Circularly-polarized waves}

As an example of circularly-polarized solutions, we consider the case 
where $\theta =0$.
In that case, we deduce from \eqref{Adot}--\eqref{Lambdadot}
the following solution to the equations of motion \eqref{mtn2},
\begin{equation} \label{CircEx}
\begin{array}{l}
u_1 = \alpha \sin \phi e^{- \omega N^0_\odot y} \cos \omega (N^0_\odot
\mathbf{m \cdot x} - t), \\ u_2 = -\alpha e^{- \omega N^0_\odot y}
 \sin \omega (N^0_\odot \mathbf{m \cdot x} - t), \\
u_3 = - \alpha \cos \phi e^{- \omega N^0_\odot y} \cos \omega
(N^0_\odot \mathbf{m \cdot x} - t)\quad\mbox{and}\quad p^* =0.
\end{array}
\end{equation}
Here $\alpha$, $\omega$ and $m$ are arbitrary, $\cos \phi$ and $\sin
\phi$ are defined by  \eqref{n+-}, $N^0_\odot$ is given by
\eqref{Ndot}--\eqref{Lambdadot} with $\theta =0$, and $\mathbf{m}$ is 
the vector defined by
\[
\mathbf{m} =  \mathbf{j} \wedge \mathbf{n^+} + \sqrt{m^2 -1}
\mathbf{n^+} = ( \sin \phi + \sqrt{m^2 -1}\cos \phi) \mathbf{i} -
(\cos \phi - \sqrt{m^2 -1}  \sin \phi)\mathbf{k}.
\]
Note that $N^0_\odot$ is real, and therefore the planes of constant
phase are orthogonal to the planes of constant amplitude
(see \eqref{S'.S''}).

The wave described by \eqref{CircEx} is circularly-polarized in the
plane orthogonal to the acoustic axis $ \mathbf{n^+}$, as would be 
the case for a homogeneous plane wave.
The wave is attenuated in the $y$-direction and travels with speed 
$(mN^0_{\odot})^{-1}$ in the direction of $\mathbf{m}$, which is not 
along an acoustic axis, in contrast to homogeneous waves.

\subsection{Elliptically-polarized waves, $\mathbf{C \cdot C} \ne 0$}
\label{Elliptically_polarized_waves_Cnonisotropic}

Now we consider elliptically-polarized waves.
In order to find the planes of polarization, we follow a procedure
introduced by Boulanger and Hayes \cite{BoHa92} for the propagation
of finite-amplitude homogeneous plane waves in a deformed Mooney--Rivlin
material.
However, their method, which dealt with real vectors, is generalized
here to the case of bivectors.

We know from section~\ref{Circularly_polarized_waves_Cnonisotropic}
that inhomogeneous plane waves 
(with a nonisotropic  $\mathbf{S}$) 
are \textit{not} circularly-polarized when the projection of the 
ellipse of $\mathbf{C}$ upon the plane orthogonal to 
$\mathbf{n}^{\pm}$ is not a circle. 
Choosing $\mathbf{a}$ and $\mathbf{b}$, two unit vectors such that 
($\mathbf{n}^{\pm}$, $\mathbf{a}$, $\mathbf{b}$) form an orthogonal 
triad, we decompose  $\mathbf{C}$ as
$\mathbf{C} = (\mathbf{C \cdot a}) \mathbf{a} 
+ (\mathbf{C \cdot b}) \mathbf{b} 
  + (\mathbf{C \cdot n^+}) \mathbf{n^+}$, and see that we must have
\begin{equation}
(\mathbf{C \cdot a})^2 + (\mathbf{C \cdot b})^2 \ne 0, 
\quad  \mathrm{or} \quad
\mathbf{C \cdot C}- (\mathbf{n^+} \cdot \mathbf{C})^2 \neq 0.
\end{equation}
This last inequality is equivalent to
\begin{equation} \label{n+-PIn+-}
\mathbf{n^{\pm} \cdot} \mbox{\boldmath $\Pi$} \mathbf{n^{\pm}} \neq 0.
\end{equation}

Within this context, we seek the nonisotropic eigenbivectors
$\mathbf{A_{\pm}}$ (say) of the tensor 
$\mbox{\boldmath $\Pi$}[C (\mathbf{C \cdot } \mathbb{B}
\mathbf{C}) \mathbf{1} + D (\mathbf{C \cdot C}) \mathbb{B}^{-1}] 
\mbox{\boldmath $\Pi$}$ such that $\mathbf{A_{\pm} \cdot C} = 0$. 
We note that with the Hamilton cyclic decomposition \eqref{hamil} of 
the tensor $\mathbb{B}^{-1}$, we may write the tensor 
$\mbox{\boldmath $\Pi$}\mathbb{B}^{-1} \mbox{\boldmath $\Pi$}$ as
\begin{equation} \label{PiBPi}
\mbox{\boldmath $\Pi$}\mathbb{B}^{-1} \mbox{\boldmath $\Pi$} 
 = \lambda_2^{-2} \mbox{\boldmath $\Pi$}
-\textstyle{\frac{1}{2}}(\lambda_3^{-2}-\lambda_1^{-2})
(\mbox{\boldmath $\Pi$}\mathbf{n^+} \otimes \mbox{\boldmath $\Pi$}\mathbf{n^-} + \mbox{\boldmath $\Pi$}\mathbf{n^-} \otimes
\mbox{\boldmath $\Pi$}\mathbf{n^+}).
\end{equation}
Using \eqref{n+-PIn+-}, we construct the bivectors $\mathbf{H^{\pm}}$,
given by
\begin{equation}
\mathbf{H^{\pm}} = 
\frac{\mbox{\boldmath $\Pi$}\mathbf{n^{\pm}}}
 {\sqrt{\mathbf{n^{\pm}\cdot}\mbox{\boldmath $\Pi$}\mathbf{n^{\pm}}}},
 \quad \mathbf{H^{\pm}  \cdot  H^{\pm}} = 1.
\end{equation}
Note that $\mbox{\boldmath $\Pi$}\mathbf{H^{\pm}}=\mathbf{H^{\pm}}$, 
so that we may write \eqref{PiBPi} of the tensor 
$\mbox{\boldmath $\Pi$}\mathbb{B}^{-1} \mbox{\boldmath $\Pi$}$ as
\[
\mbox{\boldmath $\Pi$}\mathbb{B}^{-1} \mbox{\boldmath $\Pi$}
 = \lambda_2^{-2} \mbox{\boldmath $\Pi$} 
 -\textstyle{\frac{1}{2}} (\lambda_3^{-2}-\lambda_1^{-2})
\sqrt{(\mathbf{n^+  \cdot } \mbox{\boldmath $\Pi$} \mathbf{n^+})
 (\mathbf{n^-  \cdot }\mbox{\boldmath $\Pi$}\mathbf{n^-})}
[\mathbf{H^+} \otimes \mathbf{H^-} + \mathbf{H^-} \otimes
\mathbf{H^+}].
\]

  From the definition \eqref{n+-} of $\mathbf{n}^{\pm}$ and 
\eqref{hamil}, we note that 
$\mathbf{n}^+ \wedge \mathbf{n}^-$ (parallel to $\mathbf{j}$), 
$\mathbf{n}^+ + \mathbf{n}^-$ (parallel to $\mathbf{i}$),  and 
$\mathbf{n}^+ - \mathbf{n}^-$ (parallel to $\mathbf{k}$) are orthogonal
eigenvectors of  $ \mathbb{B}^{-1}$. 
Similarly, it can be checked that the eigenbivectors of 
the symmetric operator 
$\mbox{\boldmath $\Pi$}\mathbb{B}^{-1} \mbox{\boldmath $\Pi$}$
are $\mathbf{C}$ (parallel to $ \mathbf{H}^+ \wedge \mathbf{H}^-$) and
the orthogonal bivectors $\mathbf{A}_{\pm}$ defined by
\begin{equation} \label{A+-}
\mathbf{A}_{\pm}
 = \mathbf{H^+} \pm \mathbf{H^-} 
= \frac{\mbox{\boldmath $\Pi$} \mathbf{n^+}}
     {\sqrt{\mathbf{n^+  \cdot } \mbox{\boldmath $\Pi$} \mathbf{n^+}}} 
\pm \frac{\mbox{\boldmath $\Pi$} \mathbf{n^-}}
  {\sqrt{\mathbf{n^-  \cdot } \mbox{\boldmath $\Pi$} \mathbf{n^-}}}.
\end{equation}
Now, because 
$\mbox{\boldmath $\Pi$}\mathbf{A}_{\pm}=\mathbf{A}_{\pm}$, 
we can find the amplitude bivectors $\mathbf{A}$, solutions to 
\eqref{mtnNiso}, and orthogonal to the 
propagation bivector $\mathbf{C}$.
They are the two bivectors $\mathbf{A} _{\pm}$ given by \eqref{A+-} and
their corresponding eigenvalues $\rho N^{-2}_{\pm}$ are given by
\eqref{rho+-}, or in this case, by
\begin{eqnarray*}
\rho N^{-2}_{\pm}&=& C (\mathbf{C \cdot } \mathbb{B} \mathbf{C})
  +\frac{D}{2}  (\mathbf{C \cdot C})
\{ (\lambda_3^{-2}+ \lambda_1^{-2})\\
  &&\mbox{} +(\lambda_3^{-2}-\lambda_1^{-2})
[\frac{(\mathbf{n^+} \cdot \mathbf{C})
(\mathbf{n^-} \cdot \mathbf{C})}{\mathbf{C \cdot C}}
\mp \sqrt{(\mathbf{n^+  \cdot } \mbox{\boldmath $\Pi$} \mathbf{n^+})
(\mathbf{n^-  \cdot }\mbox{\boldmath $\Pi$}\mathbf{n^-})}] \}.
\end{eqnarray*}
We note that
\begin{eqnarray}
\rho ( N^{-2}_+ - N^{-2}_-) & =& -D(\mathbf{C \cdot C})
(\lambda_3^{-2}
 -\lambda_1^{-2}) \sqrt{(\mathbf{n^+  \cdot} \mbox{\boldmath $\Pi$} 
 \mathbf{n^+})
(\mathbf{n^-  \cdot } \mbox{\boldmath $\Pi$} \mathbf{n^-})}, 
 \label{Ndistinct}\\
& =&  -D(\mathbf{C \cdot C})
(\lambda_3^{-2}-\lambda_1^{-2})
\sqrt{\Big{[}1
 - \frac{(\mathbf{n^+  \cdot C})^2}{\mathbf{C \cdot C}}\Big{]}
\Big{[}1
 - \frac{(\mathbf{n^-  \cdot C})^2}{\mathbf{C \cdot C}}\Big{]}}. \nonumber
\end{eqnarray}
When the bivector $\mathbf{C}$ and the complex quantities 
$N^{-1}_{\pm}$ are replaced by $\mathbf{n}$ (direction of propagation)
 and $c_{1}$, $c_{2}$ (speeds) for homogeneous waves, this equation 
reduces to
$ \rho (c_2^2-c_1^2)
= D (\lambda_3^{-2} - \lambda_1^{-2})\sin \phi^+ \sin \phi^-$
\cite{BoHa92}, where $ \phi^{\pm}$ are the angles between $\mathbf{n}$
and $ \mathbf{n}^{\pm}$.
This in turn is reminiscent of the `law of the product of the two
sines' (\textit{la loi du produit des deux sinus}), for the 
propagation of light through a biaxial crystal in classical linear 
optics, established empirically by Biot \cite{Biot18} in 1818 and 
theoretically by Fresnel \cite{Fres68} in 1821.

  From \eqref{Ndistinct} and \eqref{n+-PIn+-},
we see that here, the eigenvalues are distinct and therefore neither
$\mathbf{A}_+$ nor $\mathbf{A}_-$ is isotropic 
\cite[\S 3.2]{BoHa93}. 
This means that, as expected, the waves corresponding to
these amplitudes are elliptically-polarized \cite{Haye84} 
(and not circularly-polarized).
In this connection, it may be recalled that \cite{Dest99}
\textit{finite}-amplitude inhomogeneous plane waves of complex
exponential type can only be \textit{linearly}-polarized (and not
elliptically-polarized), when they propagate in an incompressible
elastic material.
Here this restriction is removed because we are dealing with 
\textit{small}-amplitude inhomogeneous plane waves.

Finally, using \eqref{mtnP=0} and the fact that
$\mathbf{A}_{\pm} \cdot \mathbf{A}_{\pm} \ne 0$, we can write
$\rho N^{-2}_{\pm}$ in terms of $\mathbf{C}$ and $\mathbf{A}_{\pm}$ as
\begin{equation} \label{NwithC&A}
 \rho N^{-2}_{\pm}= C(\mathbf{C \cdot } \mathbb{B} \mathbf{C})
+ D (\mathbf{C \cdot C})
\frac{\mathbf{A}_{\pm} \cdot \mathbb{B}^{-1} \mathbf{A}_{\pm}}
{\mathbf{A}_{\pm} \cdot \mathbf{A}_{\pm}}.
\end{equation}
For homogeneous plane waves, these expressions reduce to results
established by Boulanger and Hayes \cite{BoHa92}:
replacing $N^{-2}_{\pm}$ with their squared wave speeds
$c_1$, $c_2$, and the bivectors $\mathbf{C}$ and $\mathbf{A}_{\pm}$ 
by their real unit vectors $\mathbf{n}$, $\mathbf{a}$, 
and $\mathbf{b}$, respectively, the expressions are transformed into
$\rho c_1^2 = C(\mathbf{n \cdot } \mathbb{B} \mathbf{n})
+ D (\mathbf{a \cdot }\mathbb{B}^{-1} \mathbf{a})$ and
$\rho c_2^2 = C(\mathbf{n \cdot } \mathbb{B} \mathbf{n})
+ D (\mathbf{b \cdot }\mathbb{B}^{-1} \mathbf{b})$.

We now write down explicit examples of elliptically-polarized waves.

\subsubsection*{Example 3: Elliptically-polarized waves}

In this example, we present waves propagating in a principal
direction with attenuation in another principal direction.
It is seen that for one solution, the wave is
elliptically-polarized in the plane of the slowness bivector,
while for the other solution, the wave is linearly-polarized in the
direction normal to the plane of the slowness bivector.

We take $\mathbf{C}= m\mathbf{i}+i \mathbf{j}$ with $m> 1$.

For waves corresponding to the eigenvalue $\rho N^{-2}_+$, we have for
the amplitude: $ \mathbf{A}_+= \alpha (\mathbf{i} + i m \mathbf{j})$,
with eigenvalue: $ \rho N^{-2}_+= C ( m^2 \lambda_1^2 - \lambda_2^2) +
D (m^2 \lambda_2^{-2} - \lambda_1^{-2})$, so that the corresponding
solution is given, for $N_+$ real, by
\begin{equation}
\begin{array}{l} \label{NisoEx1}
u_1 = \alpha e^{- \omega N_+ y} \cos \omega (mN_+x - t), \\
u_2 = -\alpha m e^{- \omega N_+ y} \sin \omega (m N_+x - t), \\
u_3 =0 \quad\mbox{and}\quad
p^* = 0,
\end{array}
\end{equation}
where $\omega, \alpha$ and $m$ are arbitrary ($ m>1$), and
\begin{equation} \label{NisoEx1N}
N_+= \sqrt{\rho /[ C (m^2 \lambda_1^2 - \lambda_2^2)
+ D (m^2 \lambda_2^{-2} - \lambda_1^{-2})]}.
\end{equation}

For waves corresponding to the eigenvalue $\rho N^{-2}_-$, we find the
amplitude to be: $\mathbf{A}_-= \alpha \mathbf{k}$, with eigenvalue: 
$\rho N^{-2}_-= C (m^2 \lambda_1^2 - \lambda_2^2) + D (m^2 -1)
\lambda_3^{-2}$, and the corresponding solution is given, for $N_-$ 
real, by
\begin{equation}\label{NisoEx2}
u_1 =0, \quad
u_2 =0, \quad
u_3 = \alpha e^{- \omega N_- y} \cos \omega (m N_- x - t), \quad
p^* = 0,
\end{equation}
where $\omega, \alpha$ and $m$ are arbitrary ($ m> 1$), and
\begin{equation}  \label{NisoEx2N}
N_-= \sqrt{\rho /
[ C (m^2 \lambda_1^2 -\lambda_2^2)
+ D (m^2 -1) \lambda_3^{-2}]}.
\end{equation}

The waves described by \eqref{NisoEx1} and \eqref{NisoEx2} propagate in
the $x$-direction with speed $(m N_{\pm})^{-1}$, where $m$ is 
prescribed and $N_{\pm}$ is given by \eqref{NisoEx1N} and  
\eqref{NisoEx2N}, respectively.
They are attenuated in the $y$-direction. 
The wave given by \eqref{NisoEx1} is elliptically-polarized in the 
$xy$-plane (two conjugate radii of the ellipse are $\alpha \mathbf{i}$ 
and $m \alpha \mathbf{j}$); the wave given by \eqref{NisoEx2} is 
linearly-polarized in the $z$-direction.

\subsubsection*{Example 4: Elliptically-polarized waves}

Our next example is a wave solution propagating along an acoustic axis
with attenuation in the direction orthogonal to the acoustic axes.

Let $\mathbf{C}= \sqrt{2}\mathbf{n^+} +i \mathbf{j}
  =\sqrt{2}\cos \phi \mathbf{i} +i \mathbf{j}
      + \sqrt{2} \sin \phi \mathbf{k}$,
where $2\phi$ is the angle between the acoustic axes, defined by 
\eqref{n+-}.
After some calculation, we write another explicit solution to the
incremental equations of motion \eqref{mtn2} as
\begin{align} \label{NisoEx3}
u_1 &=[ -1 \pm \frac{1-2\cos2\phi}{\sqrt{\cos 4\phi}}]
\cos\phi \, e^{- \omega N_{\pm} y}
\sin \omega (\sqrt{2} N_{\pm}\mathbf{n^+}  \cdot \mathbf{x} - t),
\vspace{2mm} \nonumber\\ 
u_2 &=-\sqrt{2}[ 1 \pm \frac{\cos2\phi}{\sqrt{\cos 4\phi}} ]
  e^{- \omega N_{\pm} y}
\cos \omega (\sqrt{2} N_{\pm}\mathbf{n^+}  \cdot \mathbf{x} - t),
\vspace{2mm}  \\
u_3 &= [ -1 \mp \frac{1+2\cos2\phi}{\sqrt{\cos 4\phi}} ]
\sin\phi \, e^{- \omega N_{\pm} y}
\sin \omega (\sqrt{2} N_{\pm}\mathbf{n^+}  \cdot \mathbf{x} - t)
\quad\mbox{and}\quad
p^*= 0, \nonumber
\end{align}
where $\omega$ is arbitrary, $\phi$ and $\mathbf{n^+}$ are defined by
\eqref{n+-}, and
\begin{equation}  \label{NisoEx3N}
\rho N_{\pm}^{-2}=
C (2\lambda_2^{-4} -\lambda_2^2) +
 \textstyle{\frac{D}{2}} \{\lambda_3^{-2}+\lambda_1^{-2}
+(\lambda_3^{-2}-\lambda_1^{-2})
[2\cos 2\phi \mp \sqrt{\cos 4\phi}] \}.
\end{equation}

The waves described by \eqref{NisoEx3} propagate in the direction of
the acoustic axis $\mathbf{n^+}$ with speed $(\sqrt{2} N_{\pm})^{-1}$,
where  $N_{\pm}$ is given by  \eqref{NisoEx3N}. 
They are attenuated in the $y$-direction, and are 
elliptically-polarized. 

\subsubsection*{Example 5: Elliptically-polarized waves}

Our last examples are inhomogeneous plane waves for which the
planes of constant phase are not orthogonal to the planes of
constant amplitude.

Let $\mathbf{C}= m \mathbf{p} +i \mathbf{q} \, (m>1)$, where
\begin{equation}
\mathbf{p}=\cos \theta \mathbf{i} + \sin \theta \mathbf{k}, \quad
\mathbf{q}=\sin \theta \mathbf{i} - \cos \theta \mathbf{k}, \quad 
0 < \theta < \pi /2.
\end{equation}
The corresponding slowness bivector $\mathbf{S}=N \mathbf{C}$ has real
 and imaginary parts  given by
\begin{equation} \label{ExampleS'S''}
\mathbf{S'}= m N' \mathbf{p} - N'' \mathbf{q}, \quad
\mathbf{S''}= m N'' \mathbf{p} + N' \mathbf{q}.
\end{equation}

It is found from \eqref{A+-} that the amplitude bivectors
$\mathbf{A_{\pm}}$ are given by
\begin{equation}
\mathbf{A_+}= \alpha (m\mathbf{q} - i\mathbf{p}) , \quad \mathbf{A_-}
 = \alpha \mathbf{j}.
\end{equation}

Corresponding to $\mathbf{A_+}$, we have the inhomogeneous wave,
\begin{equation} \label{NisoEx4}
\mathbf{u}= \alpha e^{-\omega \mathbf{S''  \cdot  x}} 
 [m  \mathbf{q} \cos \omega(\mathbf{S'  \cdot  x}-t)
    + \mathbf{p} \sin \omega (\mathbf{S'  \cdot  x}-t)].
\end{equation}
The  directions of propagation and attenuation are those of
$\mathbf{S'}$ and $\mathbf{S''}$, given by \eqref{ExampleS'S''}, where
$N'$ and $N''$ are the real and imaginary parts of $N$, which is given
  from \eqref{NwithC&A} by
\begin{eqnarray*}
\rho N^{-2}&=&
C[m^2 (\mathbf{p \cdot }\mathbb{B}\mathbf{p})
   -(\mathbf{q \cdot }\mathbb{B}\mathbf{q})]
+ D {\frac{m^2 -1}{m^2 + 1}}
[m^2 (\mathbf{q \cdot }\mathbb{B}^{-1}\mathbf{q})
  -(\mathbf{p \cdot }\mathbb{B}^{-1}\mathbf{p})]\\
&&\mbox{}+ 2im[ C (\mathbf{p \cdot }\mathbb{B}\mathbf{q})
  - D {\frac{m^2 -1}{m^2 + 1}}
     (\mathbf{p \cdot }\mathbb{B}^{-1}\mathbf{q})].
\end{eqnarray*}
This wave propagates in the direction of  $\mathbf{S'}$ with speed 
$| \mathbf{S'} |^{-1}$, is attenuated in the direction of 
$\mathbf{S''}$, and is elliptically-polarized (two conjugate radii of 
the ellipse are $m \mathbf{p}$ and $\mathbf{q}$).

Corresponding to $\mathbf{A_-}$, we have the wave
\begin{equation} \label{NisoEx5}
\mathbf{u}= \alpha \mathbf{j} e^{-\omega \mathbf{S''  \cdot  x}}
\cos \omega (\mathbf{S'  \cdot  x}-t).
\end{equation}
Here, the respective directions of propagation and attenuation are
those of $\mathbf{S'}$ and $\mathbf{S''}$, given by \eqref{S'S''},
where $N'$ and $N''$ are the real and imaginary parts of $N$, which 
is given from \eqref{NwithC&A} by
\begin{equation}
\rho N^{-2}=
C[m^2 (\mathbf{p \cdot }\mathbb{B} \mathbf{p})
- (\mathbf{q \cdot }\mathbb{B} \mathbf{q})]
+ D (m^2 -1) \lambda_2^{-2}
  +2 i  m  (\mathbf{p \cdot } \mathbb{B}\mathbf{q}).
\end{equation}
This wave propagates in the direction of  $\mathbf{S'}$
with speed $| \mathbf{S'} |^{-1}$, is attenuated in the direction
 of  $\mathbf{S''}$, and is linearly-polarized in the $y$-direction.
In this connection, it may be noted that \textit{finite}-amplitude
linearly-polarized inhomogeneous plane waves of complex exponential
type can propagate in a homogeneously deformed Mooney--Rivlin material
only when the planes of constant phase (orthogonal to $ \mathbf{S'}$)
are orthogonal with the planes of constant amplitude  (orthogonal
to $ \mathbf{S''}$) \cite{Dest00}. Here, from \eqref{ExampleS'S''},
 we have $ \mathbf{S'  \cdot  S''}=(m^2 -1) N' N'' \ne 0$,
and these planes are not orthogonal to each other.
This is possible because we are dealing with \textit{small}-amplitude
inhomogeneous waves only.

\section{Summary and concluding remarks}

We have found all possible inhomogeneous small-amplitude motions of 
complex exponential type that can be superimposed upon the large 
homogeneous static deformation of a Mooney--Rivlin material.
The results are summarized in Tables 1 and 2, where the solutions of 
exponential sinusoidal form are listed for inhomogeneous plane waves 
with an isotropic and a nonisotropic bivector $\mathbf{C}$, 
respectively.
In the last column of the tables, it is recalled whether or not each
motion is possible with a finite amplitude, and the corresponding
reference is given.
In the second column, the equation numbers of corresponding examples 
in this paper are given.
Thus for instance, line 1.3 reads as follows: a small-amplitude
circularly-polarized inhomogeneous plane wave of complex exponential
type with isotropic propagation bivector $\mathbf{C}$ can propagate in
a homogeneously deformed Mooney--Rivlin material;
for this wave, the amplitude bivector $\mathbf{A}$ is isotropic and
 orthogonal to $\mathbf{C}$;
an example of such solution is given by equation \eqref{IsoEx2};
also note that $\mathbf{A}$ is parallel to $\mathbf{C}$ and that the 
complex scalar slowness $N$ is arbitrary;
finally, such a finite-amplitude wave solution is not possible in
a homogeneously deformed Mooney--Rivlin material (see \cite{Dest99}).
\begin{table}
\begin{center}
\begin{tabular}{|l|c|c|c|}
\hline
 Polarization Type
& Displacement
& Notes
& Finite case?
\\
\hline
\hline
\rule[-3mm]{0mm}{8mm}
 \textbf{1.1} Linear
& $ \{ \mathbf{A}
   e^{i \omega ( N \mathbf{C \cdot x} -t)} + \mathrm{c.c.} \}$,
&
&
\\
& $\mathbf{A \cdot C}=0$, $\mathbf{A \wedge \widetilde{A} =0}$.
&$  \mathbf{A} = \alpha_2 \mathbf{C} \wedge \mathbf{\widetilde{C}}$,
& yes \cite{Dest00}
\\
& Example: \eqref{IsoEx1a} with $\alpha_1=0$
&
&
\\
\hline
\rule[-3mm]{0mm}{8mm}
\textbf{1.2} Elliptical
& $ \{ \mathbf{A}
   e^{i \omega ( N \mathbf{C \cdot x} -t)} + \mathrm{c.c.} \}$,
&
&
\\
&$\mathbf{A \cdot C}=0$, $\mathbf{A \cdot A} \ne 0$.
& $\mathbf{A} = \alpha_1 \mathbf{C}
 + \alpha_2 \mathbf{C} \wedge \mathbf{\widetilde{C}}$
&
\\
& Example: \eqref{IsoEx1a} with $\alpha_1 \ne 0$
&
& no \cite{Dest99}
\\
\cline{1-3}
\rule[-3mm]{0mm}{8mm}
\textbf{1.3} Circular
&  $ \{ \mathbf{A}
   e^{i \omega ( N \mathbf{C \cdot x} -t)} + \mathrm{c.c.} \}$,
&
&
\\
&$\mathbf{A \cdot C}=0$, $\mathbf{A \cdot A} = 0$.
& $  \mathbf{A} = \alpha_1 \mathbf{C}$,
&
\\
& Example:  \eqref{IsoEx2}
& $N$ is arbitrary.
&
\\
\hline
\end{tabular}
\caption{Inhomogeneous plane waves,
                       $\mathbf{C}$ isotropic ($\mathbf{C \cdot C}=0$)}
\end{center}
\end{table}
\begin{table}
\begin{center}
\begin{tabular}{|l|c|c|c|}
\hline
 Polarization Type
& Displacement
& Notes
& Finite case?
\\
\hline
\hline
\rule[-3mm]{0mm}{8mm}
\textbf{2.1} Linear
& $ \{ \mathbf{A}
   e^{i \omega ( N \mathbf{C \cdot x} -t)} + \mathrm{c.c.} \}$,
& $\mathbf{A}$ along
& only when
\\
&$\mathbf{A \cdot C}=0$, $\mathbf{A \wedge \widetilde{A} = 0}$.
& principal axis
& $ \mathbf{S'  \cdot  S''}=0$,
\\
& Examples:  \eqref{NisoEx2},  \eqref{NisoEx5}
& of basic strain
& ($ \mathbf{S}=N \mathbf{C}$) \cite{Dest00}
\\
\hline
\rule[-3mm]{0mm}{8mm}
\textbf{2.2} Elliptical
& $ \{ \mathbf{A}
   e^{i \omega ( N \mathbf{C \cdot x} -t)} + \mathrm{c.c.} \}$,
&
&
\\
& $\mathbf{A \cdot C}=0$, $\mathbf{A \cdot A} \ne 0$.
& $( \mathbf{n^{\pm}  \cdot  C})^2 \ne  \mathbf{C \cdot C}$
&
\\
&   Examples: \hspace*{\fill}
&
&
\\
&\hspace*{\fill} \eqref{NisoEx1},  \eqref{NisoEx3},  \eqref{NisoEx4}
&
& no \cite{Dest99}
\\
\cline{1-3}
\rule[-3mm]{0mm}{8mm}
\textbf{2.3} Circular
& $ \{ \mathbf{A}
   e^{i \omega ( N \mathbf{C \cdot x} -t)} + \mathrm{c.c.} \}$,
&
&
\\
& $\mathbf{A \cdot C}=0$, $\mathbf{A \cdot A} = 0$.
& $ \mathbf{A \cdot  n^{\pm}} =0$
&
\\
& Example:  \eqref{CircEx}
&
&
\\
\hline
\end{tabular}
\end{center}
\caption{Inhomogeneous plane waves,
                       $\mathbf{C}$ nonisotropic ($\mathbf{C \cdot C \ne 0}$)}
\end{table}

Through the use of bivectors, \textit{all} inhomogeneous solutions 
of complex exponential type for the problem of small deformations 
superposed on a large static triaxial strain of a Mooney--Rivlin 
incompressible hyperelastic material were obtained systematically.
A great diversity and richness of solutions was uncovered, 
using the Directional Ellipse method \cite{Haye84}.

For inhomogeneous  waves with an isotropic slowness bivector 
(\S \ref{Cisotropic}), any direction of polarization or plane of 
polarization is permitted for linear or elliptical polarization, 
respectively;
also, for circular polarization the complex scalar slowness can be 
arbitrarily prescribed.

For inhomogeneous  waves with a nonisotropic slowness bivector
(\S \ref{Cnonisotropic}), the secular equation was established and 
solved (\S \ref{Secular_equation}), generalizing the secular equation 
for homogeneous waves \cite{BoHa95}.
Then it was seen that such linearly-polarized waves 
(\S \ref{Linearly_polarized_waves_Cnonisotropic}) can only propagate 
along one of the principal axes of the finite static stretch; 
that circularly-polarized waves 
(\S \ref{Circularly_polarized_waves_Cnonisotropic}) can only be 
polarized in one of the two planes of central circular section of the 
$\mathbb{B}^{-1}$-ellipsoid, but can propagate in directions other 
than the directions of the acoustic axes of this ellipsoid; 
and that all elliptically-polarized wave solutions
(\S \ref{Elliptically_polarized_waves_Cnonisotropic}) can be obtained 
by generalizing a method introduced in 
\cite{BoHa92} to the consideration of bivectors.

Also, in contrast to the case of \textit{finite}-amplitude
inhomogeneous plane waves \cite{Dest00, Dest99}, it was seen 
(\S \ref{Cisotropic} and 
\S \ref{Elliptically_polarized_waves_Cnonisotropic}) that elliptical
polarization is possible, and that for waves linearly-polarized in a 
principal direction 
(\S \ref{Linearly_polarized_waves_Cnonisotropic} and Example 5),
the planes of constant phase need not be orthogonal to the planes 
of constant amplitude.

Finally, it is of interest to note that the results established here
can easily be specialized to the case of a \textit{biaxial} 
pure homogeneous static prestrain, simply by taking 
$\lambda_1 = \lambda_2 = \lambda$
(say), and $\lambda_3 = \mu = \lambda^{-2}$ (say).
In that case, the tensor 
$\mbox{\boldmath $\Pi$} \mathbb{B}^{-1}\mbox{\boldmath $\Pi$}$ 
may be written as \cite{BoHa92}
\begin{equation} 
\mbox{\boldmath $\Pi$} \mathbb{B}^{-1}\mbox{\boldmath $\Pi$}  =
\lambda^{-2} \mbox{\boldmath $\Pi$} 
 - (\lambda^{-2} - \mu^{-2}) \mbox{\boldmath $\Pi$} \mathbf{k}
	\otimes \mbox{\boldmath $\Pi$} \mathbf{k}.
\end{equation}
It follows that the computation of the amplitude bivectors and 
corresponding complex scalar slownesses is greatly simplified,
and that the results of Boulanger and Hayes \cite{BoHa96} are directly
recovered from the analysis presented here.


\end{document}